\documentclass[prl,twocolumn,aps,ltxgridinfo,%
	showpacs,superscriptaddress]{revtex4}
\input{psfig}
\usepackage{bm}

\def\aV{a}
\def\cV{\lambda}
\def\rV{r}
\def\GreenG{D}
\def\LG{L_\Gamma}
\def\chiX{\chi_{_{\scriptstyle X}}}
\def\Gammaf{\Gamma_{\rm f}}
\def\Ef{X_{\rm f}}
\def\omir{\omega_{\rm ir}}

\def\tc{\tau_{\rm c}}

\begin{document}

\title{Dephasing of solid-state qubits at optimal points}

\author{Yuriy Makhlin}
        \affiliation{Institut f\"ur Theoretische Festk\"orperphysik,
         Universit\"at Karlsruhe, D-76128 Karlsruhe, Germany}
        \affiliation{Landau Institute for Theoretical Physics,
         Kosygin st. 2, 117940 Moscow, Russia}
\author{Alexander Shnirman}
        \affiliation{Institut f\"ur Theoretische Festk\"orperphysik,
         Universit\"at Karlsruhe, D-76128 Karlsruhe, Germany}

\begin{abstract}
Motivated by recent experiments with Josephson-junction circuits, we analyze
the influence of various noise sources on the dynamics of two-level systems
at optimal operation points where the linear coupling to low-frequency 
fluctuations is suppressed.
We study the decoherence due to nonlinear (quadratic) coupling,
focusing on the experimentally relevant $1/f$ and Ohmic noise power spectra. For
$1/f$ noise strong higher-order effects influence the evolution.
\end{abstract}

\pacs{85.25.Cp, 03.65.Yz, 75.10.Jm}

\maketitle

For quantum-information processing it is crucial to preserve phase coherence.
Recent experiments~\cite{Nakamura_Nature+2qb, Saclay_Manipulation_Science,
Han_Manipulation_Science, Martinis_Rabi_PRL, Delft_Rabi} with Josephson-junction 
circuits demonstrated long-lived 
coherent oscillations. They showed resolution, sufficient for detailed studies
of the dephasing times and decay laws, stressing the need for the theory 
analysis of the dissipative dynamics of qubits subject to relevant noise 
sources.

In solid-state systems decoherence is potentially strong due the host of
microscopic modes. In Josephson qubits the noise is dominated by
material-dependent sources, such as background-charge fluctuations or variations
of critical currents and magnetic fields, with power spectrum peaked at low
frequencies, often $1/f$.
A further relevant contribution is the electromagnetic noise of
the control circuit, typically Ohmic at low frequencies.
The $1/f$ noise appears difficult to suppress and, since the dephasing is 
dominated by low-frequency noise, it is particularly destructive.
On the other hand, \textcite{Saclay_Manipulation_Science} showed
that the effect of this noise can be substantially reduced by tuning
the linear longitudinal qubit-noise coupling to zero. The same strategy, 
suppressing the linear qubit-detector coupling, was used to minimize the effect 
of the quantum detector in the off-state. The achieved coherence time was 2--3 
orders of magnitude longer than in earlier experiments.

The $1/f$ noise (more generally, strong low-frequency noise) plays a major
role in many solid-state systems. The long-range correlations in time make the 
analysis of its effect difficult. In this letter we analyze the 
dynamics of a qubit subject to singular noise, with a focus on nonlinear 
coupling and higher-order effects. We also apply the developed formalism to an
environment with regular (e.g., Ohmic thermal) noise.

The dynamics of a dissipative two-level system (spin 1/2, qubit) can be
described by the Hamiltonian:
\begin{equation}
{\cal H}=
-\frac{1}{2}\left(\varepsilon\;\sigma_z +V\;\sigma_z + U\;\sigma_x\right)
+ H_{\rm bath}
\,.
\label{Eq:HamV}
\end{equation}
The longitudinal ($V$) and transverse ($U$) fluctuations may result
from various microscopic noise sources, described by $H_{\rm bath}$.
In this letter we analyze the dephasing in the situation of quadratic 
longitudinal coupling,
\begin{equation}
V = \cV X^2\,,\qquad U=0\,,
\label{Eq:V=X2}
\end{equation}
to a source of Gaussian noise $X(t)$, with noise power $S_X(\omega)$. 
Here $X(t)$ is a basic physical quantity (e.g., voltage or magnetic field),
which controls the qubit's Hamiltonian~\footnote{The analysis is easily 
generalized to several noise sources coupled to arbitrary spin components.}. 
This model is relevant to a Josephson qubit at an optimal point, 
investigated in recent experiments~\cite{Saclay_Manipulation_Science}.
To emphasize specific features of this model, we first
recall the description of the dissipative qubit dynamics when the noise is 
either short-correlated or Gaussian longitudinal. Then we motivate 
model~(\ref{Eq:V=X2}) and show that it can describe the effect of both 
longitudinal and transverse low-frequency noise. We discuss the statistics of 
$X$ and $V$, and then analyze dephasing in model~(\ref{Eq:V=X2}). 
   
For weak short-correlated noise the dynamics is described by the Bloch
equations, regardless of the noise statistics.  The weak dissipative effects
from many uncorrelated time intervals $\sim\tc$ (the correlation time)
accumulate and the effect of both longitudinal and transverse noise can be
described by the markovian Bloch equations.  The dissipative rates are given by
the golden rule:  the $\sigma_z$-relaxation rate $1/T_1 =
S_U(\omega=\varepsilon)/2$ and the rate of dephasing (decay of $\sigma_{x,y}$)
$1/T_2 = 1/(2T_1)+1/T_2^*$, the `pure' dephasing rate $1/T_2^* =
S_V(\omega=0)/2$ being dominated by low $\omega$.  This approach applies for
weak noise with a short correlation time $\tc\ll T_1, T_2$.

One can still rely on the lowest order of the cumulant expansion (but beyond the 
golden rule) for Gaussian longitudinal noise $V$ ($U=0$), even for long 
correlations. The coherence $\langle \sigma_{-}(t)\rangle$ (here 
$\sigma_-=(\sigma_x-i\sigma_y)/2$) 
decays then as     
\begin{equation}
\label{Eq:GR}
|\langle \sigma_{-}(t)\rangle| =
\exp\left(
-\frac{1}{2}\int\,\frac{d\omega}{2\pi}\, S_V(\omega) \,
\frac{\sin^2(\omega t/2)}{(\omega/2)^2}
\right)
\,,
\end{equation}
where $S_V = \frac{1}{2} \langle\langle [V(t),V(0)]_+ 
\rangle\rangle_\omega$ is the noise power.
For instance, for a linear coupling, $V(X)=\aV X$, to a Gaussian-distributed 
$1/f$ (flicker) noise $X$, when $T_2^*$ defined above vanishes, one finds the 
dephasing law $\exp(-\aV^2\Ef^2t^2|\ln(\omir t)|/2\pi)$ 
(cf.~Refs.~\onlinecite{Cottet_Naples,Nakamura_Echo}). Here $\Ef^2$ sets the 
magnitude of the noise,
\begin{equation}
S_X(\omega)=\Ef^2/|\omega| \,,
\end{equation}
and may depend on the external conditions, such as temperature.  
The infra-red cutoff $\omega_{\rm ir}$ may be set, and controlled, by the 
details of an 
experiment. For instance, when a measurement of dephasing, performed over a 
short time $t$, 
is averaged over many runs, the fluctuations with frequencies down to the 
inverse of the total signal acquisition time contribute to the phase 
randomization~\cite{Cottet_Naples}
(this averaging improves the accuracy and is needed to monitor the oscillations 
of expectation values~\cite{Nakamura_Nature+2qb, Saclay_Manipulation_Science,
Han_Manipulation_Science, Martinis_Rabi_PRL, Delft_Rabi}). In an echo-type 
experiment the contribution of slow 
fluctuations can be suppressed~\cite{Nakamura_Echo}. The controlled frequency of 
the compensating echo pulses sets $\omega_{\rm ir}$ (in 
Ref.~\onlinecite{Nakamura_Echo} $\omega_{\rm ir}\sim 1/t$ but it can be varied). 
Here we focus on the case $\omega_{\rm ir} t \gtrsim 1$.

Typically, in the mentioned regime of linear coupling to $1/f$ noise dephasing 
is strong. To increase the coherence time one
tunes the system to a point where the linear term vanishes, 
$\aV=0$~\cite{Saclay_Manipulation_Science}. The
effect of the remaining quadratic coupling is described by the model
(\ref{Eq:V=X2}). Moreover, this model can also account for low-frequency 
transverse fluctuations $U_{\rm lf}$. Indeed, in the 
adiabatic approximation we diagonalize (\ref{Eq:HamV}) to 
$-\frac{1}{2}\sqrt{(\varepsilon+V)^2 + U_{\rm
lf}^2}\;\sigma_z \approx -\frac{1}{2}(\varepsilon + V_{\rm eff})\;\sigma_z$, 
where $V_{\rm eff} = V + U_{\rm lf}^2/(2\varepsilon)$. Hence as long as the 
relaxation due to the resonant ($\omega\sim\varepsilon$) part of $U$ is 
negligible, it is sufficient to analyze the model (\ref{Eq:V=X2}). 

This analysis requires an account of higher orders, and thus the
knowledge not only of $S_V(\omega)$ but of the full statistics of $V$. The 
latter depends on the statistics of the basic quantity $X$
as well as on the qubit-noise coupling.  In this letter we consider a
Gaussian-distributed $X$ (see below), but due to the nonlinear coupling in
Eq.~(\ref{Eq:V=X2}) the qubit is subject to non-Gaussian fluctuations $V=\lambda
X^2$.  In combination with long correlations of $1/f$ noise, this shows that
further analysis is needed.

The statistics of the low-frequency fluctuations of the basic quantity $X$
deserves further discussion.  The assumed Gaussian statistics is generic for
noise produced by many microscopic modes, due to the central limit theorem and
regardless of the noise mechanism.  As for the flicker (`$1/f$') noise in
mesoscopic circuits, it may be dominated by only a few modes (bistable
fluctuators) or contain comparable contributions of many of those, depending on
the sample, and one may expect non-Gaussian resp.  Gaussian noise.  This was
demonstrated in experiments at kHz- and lower frequencies~\cite{flicker-low}.
While the noise at higher frequencies MHz--GHz, relevant for the dephasing of
qubits, is less explored, recent data suggest that it may have the same
nature~\cite{Nakamura_Echo,flicker-high,Delft_Rabi}.  Here we focus on the
analysis of the influence of Gaussian noise $X$; the effect of a few bistable
systems was discussed recently in Ref.~\onlinecite{Paladino_1/f}.  Our analysis
on one hand, indicates interesting features of decoherence for quadratic
coupling; on the other hand, one can consider the qubit as a probe of the noise,
thus our findings may help in identifying the noise mechanism.

{\it Analysis of dephasing} in model (\ref{Eq:V=X2}).
To be definite, we assume that $\cV>0$; the
sign change has no effect on the dephasing laws, but reverses the phase
accumulated due to the nonzero average of $V$.
We follow the evolution of the off-diagonal entry of 
the qubit's density matrix $\langle\sigma_-(t)\rangle = \langle S^\dagger 
\sigma_- S\rangle$, with the 
evolution operator $S=T\exp(-\frac{i}{2}\int_0^t V\sigma_z dt')$, which yields
$\langle\sigma_{-}(t)\rangle = P(t) e^{i\varepsilon t/\hbar} 
\langle\sigma_{-}(0)\rangle$,
where
\begin{equation}
P(t) = \left\langle
\tilde T\!\exp\left({\frac{i}{2}\int\nolimits_{0}^{t} V dt'}\right)
T\!\exp\left({\frac{i}{2}\int\nolimits_{0}^{t} V dt'}\right)
\right\rangle
\,,
\label{Eq:s+}
\end{equation}
with averaging over noise realizations. $T$ and $\tilde T$ denote time resp.
reverse-time ordering; their combination in  Eq.~(\ref{Eq:s+}) 
corresponds to the Keldysh-time ordering. Note the same signs in $\tilde T\exp$ 
and $T\exp$, selected by the operator $\sigma_-$.

{\it Gaussian approximation\/}.
In the lowest-order perturbative analysis one can use 
Eq.~(\ref{Eq:GR})~\cite{Shnirman_Makhlin_Schoen_Nobel}.
For Ohmic fluctuations of $X$, with noise power $S_X(\omega) = \rV \omega 
\coth(\omega/2T)$, one finds $S_{X^2}(\omega) = (\rV^2/3\pi) \omega 
(\omega^2+4\pi^2T^2) \coth(\omega/2T)$ and Eq.~(\ref{Eq:GR}) yields for weak 
noise $\cV\rV T\ll 1$ the exponential decay with rate
\begin{equation}
\frac{1}{T_2^*} = \frac{\cV^2}{2} S_{X^2}(\omega=0) 
=\frac{4\pi}{3}(\cV \rV)^2 T^3
\,.
\label{Eq:Tcube}
\end{equation}
This dephasing is stronger suppressed by cooling compared to the case of linear 
coupling, when $1/T_2^*\sim T$.

For $1/f$ noise one finds $S_{X^2}= (4/\pi)\Ef^4 
\ln|\omega/\omega_{\rm ir}| / 
|\omega|$ and $P(t)= \exp(-[\Gammaf t \ln (\omega_{\rm ir}t)/\pi]^2)$, where
\begin{equation}
\Gammaf = \cV {\Ef^2}
\,.
\label{Eq:Gamma_1/f}
\end{equation}

{\it Higher orders: results and discussion\/}. 
Below we analyze effects beyond this Gaussian  approximation.
For the Ohmic noise we confirm that Eq.~(\ref{Eq:GR}) holds at all relevant 
times,
yielding the decay rate (\ref{Eq:Tcube}). In contrast, for the $1/f$ noise 
corrections are strong:
\begin{eqnarray}
\label{Eq:decay_law_short}
|P(t)| =&
\left[1+\left(\frac{2}{\pi}\Gammaf t \ln\frac{1}{\omega_{\rm ir}t}
\right)^2\right]^{-1/4},
&\quad \Gammaf t\ll1
\,,\\
=& e^{-\Gammaf t/2}, &\quad \Gammaf t\gg1
\,.
\label{Eq:decay_law_long}
\end{eqnarray}

\begin{figure}
\centerline{\hbox{\psfig{figure=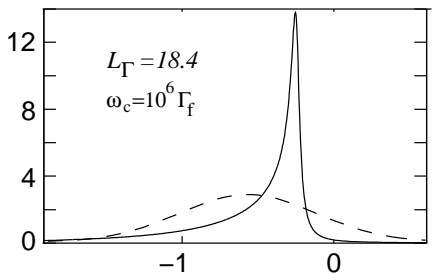,width=0.66\columnwidth}}}
\caption[]{\label{Figure:Line} Line shape $\chi_{\sigma_x}''(\omega)$, in units
of $1/(\Gammaf\LG)$ and as a function of $(\omega-\varepsilon)/(\Gammaf\LG)$.
Eq.~(\ref{Eq:decay_law_short}) sets a singular universal shape,
$\propto(\omega-\varepsilon)^{-1/2}$. The high-frequency
contribution (\ref{Eq:decay_law_long}, \ref{Eq:1/f hf})
washes out this singularity on scale $\Gammaf$ ($\LG^{-1}$ in these
units) and shifts the peak. The dashed line shows the result of the Gaussian 
approximation $F_2$ combined with the phase shift $F_1$.}
\end{figure}

At short times we find a very slow decay with time scale
$(\Gammaf\LG)^{-1}$, where $\LG\equiv \ln (\Gammaf/\omega_{\rm ir})$ and 
we assume $\LG\gg1$ (our qualitative results persist down to $\LG\sim1$): At 
very short times it reduces to $1-[\Gammaf 
t\ln(\omega_{\rm ir}t)/\pi]^2$ and coincides with the result of the Gaussian
approximation. This slow initial decay may be advantageous for application of 
quantum error correction. At longer times $t\sim(\Gammaf\LG)^{-1}$ the decay 
crosses over to a power law $\propto 1/\sqrt{t}$. Finally, at the 
(parametrically longer) time $t\sim\Gammaf^{-1}$ the decay becomes exponential 
(\ref{Eq:decay_law_long}), due to the high-frequency contribution.
For large $\LG\gg1$ a substantial decay occurs already in the range of 
Eq.~(\ref{Eq:decay_law_short}).

This unusual decay law translates into a peculiar line shape of the transverse 
spin susceptibility $\chi_{\sigma_x}''(\omega) = P(\omega-\varepsilon)/2$ (at 
$T\ll \varepsilon$), shown in Fig.~\ref{Figure:Line}: 
Eq.~(\ref{Eq:decay_law_short}) gives a singular peak 
$\propto(\omega-\varepsilon)^{-1/2}$, and the term (\ref{Eq:decay_law_long}) 
washes it out on the scale $\Gammaf$, setting the peak height 
$\sim1/\sqrt{\Gammaf}$. Apart from the dephasing, we find a phase
contribution which shifts the peak by $\Gammaf\ln(\omega_c/\Gammaf)/\pi$ (if
this logarithm exceeds $\ln\LG$; $\omega_c$ is the unltraviolet cutoff).

\begin{figure}
\centerline{%
\hbox{\psfig{figure=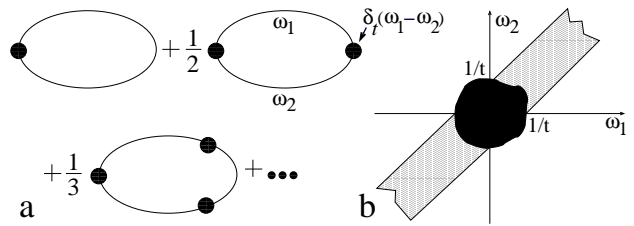,width=0.95\columnwidth}}}
\caption[]{{\bf a}.\label{Figure:Linked_Cluster} Linked-cluster expansion. The 
factor 
$\delta_t(\Delta\omega)=\sin(\Delta\omega/2)/(\pi\Delta\omega)$ at each vertex 
violates the frequency conservation.
{\bf b}.\label{Figure:Shar_Truba} `High'- and `low'-frequency regions 
dominating the integration in the cluster diagrams.}
\end{figure}

{\it Derivation\/}.
The Keldysh-time-ordered exponent (\ref{Eq:s+}) may be expanded into
the linked-cluster series:
\begin{equation}
\label{Eq:Linked_Cluster}
P(t) =\exp\left( \sum_{n=1}^{\infty}\frac{1}{n}\;F_n \right)
\,,
\end{equation}
$F_n$ representing the contribution of all connected diagrams of
order $n$ in the perturbation $V$ (the Gaussian approximation (\ref{Eq:GR})
neglects $F_{n>2}$). For the model (\ref{Eq:V=X2}) they are visualized in
Fig.~\ref{Figure:Linked_Cluster}a, with a single cluster in each order.
Here the solid lines represent the bare $2\times2$ Keldysh Green functions
of the bath $\hat D= -i \langle T_{\rm K} X(t)X(0) \rangle$, whose Keldysh and
retarded components~\cite{Rammer_Smith,Kamenev_Lectures} are related to the
noise power and the response function of
the bath, respectively:
$D^{\rm K}=-2iS_X$, $D^{\rm R}=-\chiX$.
Each vertex contributes a factor
$\cV$ and integration over the time  interval $(0;t)$.
Thus we find in the $n$-th order:
\begin{eqnarray}
\label{Eq:F_n_no_Golden_Rule}
F_n(t) =&&\frac{(-\cV)^n}{2}\,{\rm tr}
\int\nolimits_{0}^{t} dt_1
\int\nolimits_{0}^{t} dt_2 \dots
\nonumber\\
&&\hat\GreenG(t_1-t_2)\,\hat 1\,\hat\GreenG(t_2-t_3)\,\hat 1 \dots
\nonumber\\
=&&\frac{(-\cV)^n}{2}\,{\rm tr}
\int d\omega_1 d\omega_2 \dots
\nonumber\\
&&
\hat\GreenG(\omega_1) \delta_t(\omega_1\!-\!\omega_2) \hat\GreenG(\omega_2) 
\delta_t(\omega_2\!-\!\omega_3)
\dots,
\end{eqnarray}
where $\delta_t(\omega)\equiv\sin(\omega t/2)/(\pi\omega)$.
The vertices contribute the identity matrices $\hat 1$ in the Keldysh space, 
rather than the 
familiar $\hat\tau_z$, as a result of the sign structure in Eq.~(\ref{Eq:s+});
such a perturbation is called `quantum'~\cite{Kamenev_Lectures}.
For the coupling (\ref{Eq:V=X2}) the averaging in (\ref{Eq:s+}) reduces to a 
Gaussian integral and thus to a determinant of an integral operator. From this 
standpoint our calculation may be viewed as determinant regularization.

{\it Tractable regimes\/}.  The series (\ref{Eq:Linked_Cluster}),
(\ref{Eq:F_n_no_Golden_Rule}) should be evaluated for each particular time $t$.
In general, it is difficult due to its complex structure: the
propagators $\hat D$ are diagonal in frequency and the
vertices are diagonal in time.  However, one can evaluate the 
dephasing in certain limits: (a) at long times (the exact condition to be 
specified) $\delta_t\approx\delta$, 
and frequency is conserved at the vertices; (b) at short times the 
lines become time-indepednent, $\hat D(t_i-t_{i+1})\approx \hat D(\Delta t=0)$.  
These ideas are used below for the case of $1/f$ noise, when $\hat D(\Delta 
t\!=\!0)$ diverges, to find the short- and long-time behavior of $P(t)$.

{\it $1/f$ noise: (a) higher frequencies\/}.
For the $1/f$ noise it is technically more convenient to discuss the
contributions of two frequency ranges at given $t$ (see
Fig.~\ref{Figure:Shar_Truba}b), rather than the short- and long-time limits.

We begin with higher frequencies: In the integral 
(\ref{Eq:F_n_no_Golden_Rule}) the adjacent frequencies may differ by $\sim 1/t$ 
due to the vertex factors $\delta_t$. In the ($t$-dependent) range $|\omega|\gg 
1/t$ such a shift does not 
change much the propagators (noise power), and $\delta_t(\Delta\omega) \approx 
\delta(\Delta\omega)$ at each vertex. Thus we find the contribution of high 
frequencies:
\begin{equation}
\label{Eq:1/f hf}
\ln P^{\rm hf}(t)=
-t\, \int\nolimits_{\sim1/t}^\infty \frac{d\omega}{2\pi}\,
\ln\left(1-2i \lambda S_X(\omega)\right)
\,.
\end{equation}
The contribution of frequencies of order $1/t$ is only estimated by this 
expression. However,
at long times $\Gammaf t\gg 1$, when Eq.~(\ref{Eq:1/f hf}) dominates $P(t)$
(as we find below), this contribution is negligible, and we obtain 
Eq.~(\ref{Eq:decay_law_long}): $\ln|P^{\rm hf}(t)|=-\Gamma_\infty t$, where
\begin{equation}
\Gamma_\infty = \Gammaf \int\nolimits_0^\infty \frac{dx}{2\pi} \ln
\left(1+\frac{1}{x^2}\right)
= \frac{1}{2}\,\Gammaf
\,.
\label{Eq:Phf}
\end{equation}
As for the phase shift, to the logarithmic accuracy in the limit $\ln(\Gammaf 
t)\gg 1$ we find ${\rm Im}\,\ln P^{\rm hf}(t) = \Gammaf t\ln(\Gammaf t)/\pi$.


{\it $1/f$ noise: (b) lower frequencies\/}.
For the analysis of the contribution of low frequencies $|\omega|\ll 1/t$ one
may replace the vertex factor $\delta_t(\Delta\omega)$ by its value, $t/2\pi$, 
at $\Delta\omega=0$, and the correlations between frequencies of adjacent lines 
are irrelevant. Thus the series (\ref{Eq:Linked_Cluster}) gives~\footnote{The 
derivation of Eqs.~(\ref{Eq:1/f hf}), 
(\ref{Eq:1/f lf}) may involve summation of divergent series.
Other methods, e.g., based on functional integration, may be used to justify 
these results.}:
\begin{eqnarray}
\label{Eq:F_n_lf}
&\displaystyle
F_n^{\rm lf} =
\!\!\frac{1}{2}\!\left(\frac{2it}{\pi}
\int_{\omega_{\rm ir}}^{\sim1/t}\frac{\Gammaf d\omega}{|\omega|}
\right)^n
\!\!\!= \frac{1}{2}\!\left(\frac{2}{\pi}\,it\Gammaf
\ln \frac{1}{\omega_{\rm ir}t}\right)^n
\!\!\!\!\!,&
\\
\label{Eq:1/f lf}
\label{Eq:Plf}
&\displaystyle
\ln P^{\rm lf}(t) = -
\frac{1}{2}\,\ln\left(1-\frac{2}{\pi}\,it\Gammaf
\ln \frac{1}{\omega_{\rm ir}t}\right)
\,.&
\end{eqnarray}
Again, the contribution of frequencies close to the upper limit, $\omega\sim 
1/t$, is not given reliably 
by Eq.~(\ref{Eq:1/f lf}), but it is negligible to the logarithmic accuracy at 
$\ln(1/\omega_{\rm ir}t) \gg 1$.

The low and high (and intermediate) frequencies contribute at all
times, but at short times $\ln |P(t)|$ is dominated by low frequencies and at
long times by high frequencies. These leading terms yield 
Eqs.~(\ref{Eq:decay_law_short}), (\ref{Eq:decay_law_long}). (The susceptibility 
$\chi_X(\omega)$ enters the analysis, but in the final results we assumed a 
regular and hence negligible $\chi$.)

{\it Ohmic noise\/}.
Let us apply the developed approach to equilibrium thermal noise with quadratic 
longitudinal coupling, considering the example of an Ohmic bath. Its 
low-frequency spectral density $\chiX''(\omega) =
i (D^{\rm R}-D^{\rm A})/2 = \rV\omega$ is related by the fluctuation-dissipation 
theorem to the noise power: 
$D^{\rm K}=(D^{\rm R}-D^{\rm A})\coth(\omega/2T)$. The variation scale of the 
latter is set by temperature, $1/\tc\sim T$. Hence at times $t\gg 1/T$ 
changing the frequency by $\sim 1/t$ has little effect on $D^{\rm K}$, and one 
may use the long-time approach, with frequency conservation in the diagrams 
(analysis of $D^{\rm R/A}$ does not change this conclusion).

We consider the case of weak-coupling and begin with the lowest order $F_2$. 
Evaluation of this Gaussian cluster involves $\mathop{\rm tr}\hat D^2 = (D^{\rm 
R})^2 + (D^{\rm A})^2 +(D^{\rm R}-D^{\rm A})^2 [\coth^2 (\omega/2T)-1]$. The 
frequency integral  of the first two
terms (analytic in the upper/lower half-plane) vanishes, and we find exponential 
decay with rate (\ref{Eq:Tcube}) at $t\gg 1/T$ (incl. $t \sim T_2^*$).

As expected for weak noise, the higher orders provide only a small correction to 
Eq.~(\ref{Eq:Tcube}), of order 
$(\cV\rV\omega_{\rm c})^2$.  Indeed, only the even orders contribute to 
dephasing, for which $\mathop{\rm tr}\hat D^{2k}=(D^{\rm R})^{2k} + (D^{\rm 
A})^{2k} +$ terms localized at $\omega\lesssim T$. One finds that it is 
sufficient to evaluate the second-order contribution provided $\cV\hat 
D(\omega=0) \ll 1$. In the Ohmic case this weak-coupling condition reads $\cV\rV 
T, \cV\rV\omega_{\rm c}\ll 1$. 

{\it Higher (non-Gaussian) orders as screening\/}.
The exponential decay law (\ref{Eq:decay_law_long}) appears surprising for $1/f$ 
noise, with long-time correlations. In fact, it develops due to the screening of 
the long-time (low-$\omega$) fluctuations, interacting via the term 
(\ref{Eq:V=X2}), similar to the screening of interaction in the Coulomb gas.
Moreover, the calculation of $P(t)$ parallels that of the correlation energy of
the Coulomb gas~\cite{Corr_Energy}. The result can be found from the
lowest-order cluster $F_2$ with the solid lines replaced by the properly 
renormalized
(screened) Keldysh propagators $\cal D$.  Diagrammatically this appears natural:
the diagrams $F_{n>2}$ in Fig.~\ref{Figure:Linked_Cluster}a may be viewed as
$F_2$ with additional vertices on the lines.

As a result of the screening, qualitatively, the $1/f$ divergence is cut off at
$\omega\sim\Gammaf$, and the white low-$\omega$ noise produces the exponential
decay.  We omit the detailed discussion but mention a specific property of the
screened $\cal D$:  unlike the bare $D$ it is $t$-dependent (due to the
$t$-dependent vertices).  In other words, the screening sets in gradually, with 
the mentioned saturation at $t\gtrsim\Gammaf^{-1}$.

{\it Preparation effects\/}.
So far we worked under the assumption  
that at $t=0$ the bath and spin were disentangled and the bath was 
prepared in the thermal state of $H_{\rm bath}$. To see, if it can change
our conclusions, we considered a typical experiment, in which one monitors the 
decay of a superposition of the spin states in the presence of permanent 
spin-bath coupling. We found that for $1/f$ noise with a non-singular 
low-frequency susceptibility 
$\chi(\omega)$ our conculsions about the dephasing persist.

In conclusion, we have analyzed the decay laws of coherence of a qubit coupled
quadratically to the environment.  We have shown that higher-order effects
become important for certain noise spectra (notably, for $1/f$), and found the
dephasing times (\ref{Eq:Tcube})--(\ref{Eq:decay_law_long}) in various regimes.
We are grateful to G.~Sch\"on for valuable contributions to this work. We also 
thank A.~Mirlin, K.~Gaw\c{e}dzki, D.~Esteve, D.~Vion, C.~Urbina, L.~Levitov,
P.~Degiovanni, Y.~Nakamura for fruitful discussions.  The work is part of the EU
IST Project SQUBIT, the CFN of the DFG, and a 
research network of the
Landesstiftung BW.  Y.M.  was supported by the
Humboldt foundation, the BMBF, and the ZIP programme of the German government.

\bibliography{flicker}

\end{document}